# Structural transition states explored with minimalist coarse grained models: applications to Calmodulin

Francesco Delfino[1,2,*], Yuri Porozov[1,4], Eugene Stepanov[3,5], Gaik Tamazian[6], Valentina Tozzini[2]

[1] I.M. Sechenov First Moscow State Medical University, Trubetskaya st. 8-2, Moscow 119991, Russia

[2] Istituto Nanoscienze – CNR and NEST-Scuola Normale Superiore, Piazza San Silvestro 12, 56127 Pisa, Italy

[3] St Petersburg Branch of the Steklov Mathematical Institute of the Russian Academy of Sciences, Fontanka 27, 191023 St Petersburg, Russia

[4] ITMO University, 49 Kronverksky Av., St. Petersburg 197101, Russia

[5] Department of Mathematical Physics, Faculty of Mathematics and Mechanics, St Petersburg State University, Universitetskij pr. 28, Old Peterhof, 198504 St Petersburg, Russia

[6] Theodosius Dobzhansky Center for Genome Bioinformatics, 41A, Sredniy Av., St. Petersburg State University, St. Petersburg, Russia

**\* Correspondence:**
Francesco Delfino delfinofrancesco90@gmail.com

**Abstract**

Transitions between different conformational states are ubiquitous in proteins, being involved in signaling, catalysis and other fundamental activities in cells. However, modeling those processes is extremely difficult, due to the need of efficiently exploring a vast conformational space in order to seek for the actual transition path for systems whose complexity is already high in the "steady" states. Here we report a strategy that simplifies this task attacking the complexity on several sides. We first apply a minimalist coarse-grained model to Calmodulin, based on an empirical force field with a partial structural bias, to explore the transition paths between the apo- closed state and the Ca-bound open state of the protein. We then select representative structures along the trajectory based on a structural clustering algorithm and build a cleaned-up trajectory with them. We finally compare this trajectory with that produced by the online tool MinActionPath, by minimizing the action integral using a harmonic network model, and with that obtained by the PROMPT morphing method, based on an optimal mass transportation-type approach including physical constraints. The comparison is performed both on the structural and energetic level, using the coarse-grained and the atomistic force fields upon reconstruction. Our analysis indicates that this method returns trajectories capable of exploring intermediate states with physical meaning, retaining a very low computational cost, which can allow systematic and extensive exploration of the multi-stable proteins transition pathways.

## 1  Introduction

Signaling is a core activity in cells. Most of the signaling processes are regulated by bi- (or multi-) stable proteins, which can undergo conformational transitions in response to changes in environmental conditions or stimuli of different origin[1]. This class includes among others, G-proteins coupled receptors[2] such as Rhodopsins[3] and other transducers, e.g. Calmodulin[4], and a vast number of enzymes undergoing conformational changes during their activity, such as the HIV-1 protease[5]. The structural variations are usually quite large, therefore atomistic molecular dynamics (MD) simulations might not be the most proper method to address them, because the slow transition kinetics requires simulations exceeding the currently reachable time and space scales. In addition, the atomistic representation with standard force fields (FF) is not warranty of accuracy for the strongly distorted and out of equilibrium transition states[6].

Strategies to overcome these difficulties involve different actions. On one side, adopting simplified low-resolution descriptions of the system such as coarse-grained (CG) models[7] reduces the computational cost and allows performing more efficient sampling of the conformational space. This advantage comes at the cost of increasing the empirical content of the FF, and consequently reducing predictive power and transferability. A compromise between accuracy and predictive power[8] is reached by including some *a priori* knowledge of the system, in different forms, such as, e.g., a (partial) bias[9,10] towards reference structures, which is not a big restrain when one has to sample the path connecting two given structures.

On the other side, one can act by simplifying the sampling algorithm, e.g. using morphing related methods[11,12,13] without relying on any specific FF. In particular PROMPT[12,13] employs an approach based on the optimal mass transportation problem including physical constraints of geometric nature[14]. Methods based on the action minimization of simplified FFs, such as MinActionPath[15], can be thought as located between the two approaches.

In this work, we first apply a minimalist CG model for proteins to the test case of Calmodulin, chosen because of its large conformational transition upon calcium binding. We perform molecular dynamics simulations in different conditions to sample the transition path. We then compare these results with those of the simplified path sampling methods.

## 2  System and methods

### 2.1 The coarse grained model

The coarse graining procedure we consider in this work is schematized in Fig. 1(a,b), reporting the atomistic representation of a protein chain and the minimalist CG (MCG) representation in which only the Cα atoms are present. The choice of Cα as the representative atom of the amino-acid bead allows uniquely representing the secondary structure by the internal variables α,θ[16]. The interactions are described by an empirical FF, derived from an energy potential $U$ with a form similar to the atomistic ones, separated in bonded and non-bonded interactions

$$U = \sum_{\text{bonds}} u_i^b(d_i) + \sum_{\text{bond angles}} u_i^\theta(\theta_i) + \sum_{\text{dihedrals}} u_i^\phi(\phi_i) + \sum_{i>j} u^{nb}(r_{ij}) \qquad (1)$$

$d_i$, $\theta_i$, $\phi_i$ being the bond distances, angles and dihedrals describing the local geometry of connected beads and $r_{ij}$ distances between non-bonded ones (see Fig. 1b). The functional forms (reported in Table 1) are somewhat more complex than those used in atomistic FFs: while $u_i^b$ are holonomic restrains, the $u_i^\theta$ and $u_i^\phi$ take forms accounting for the anharmonicity of the CG interactions; in addition, the parameters are chosen to account for the different geometrical stiffness of the secondary structures, assigning different values to helices and sheets (see Table 1[17]). The non-bonded interactions occur between couples not already involved in a bond, bond angle or dihedral interaction and are separated in local and non-local part

$$\sum_{i>j} u^{nb}(r_{ij}) = \sum_{i,j|r_{ij}<r_{cut}} u_{loc}(r_{ij}) + \sum_{i,j|r_{ij}>r_{cut}} u_{nl}(r_{ij}) \qquad (2)$$

both represented by a Morse potential, with the local term retaining a bias towards a reference structure (see Table 1). In this work the local/non-local separation is based on a geometric criterion: all the non-bonded couples whose distance is less than $r_{cut}=8.5$Å in the reference structure are considered local, the others are considered non-local. The cutoff value used here was previously shown to include all the relevant H-bonds and other possible specific interactions such as disulfide or salt bridges[18]. The parameters of the Morse potential, were optimized in our previous works including a dependence on $r_0$ (distance in the reference structure) in order to reproduce stronger interaction in the H-bonding range and weaker ones in the hydrophobic range[19] (see Table 1). Since here we are not interested in the accurate simulation of the inter-protein interactions, the non-local part is represented by a generic amino-acid independent potential reproducing an average level of hydrophobicity (Table 1), instead than with a complex matrix of amino-acid dependent potentials[20].

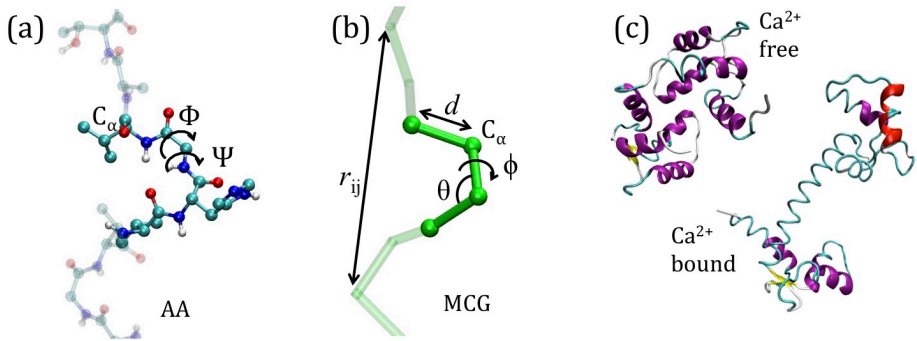

**Fig. 1.** The model system. (a) the atomistic representation of the protein chain (side chains are omitted for clarity) (b) coarse grained representation. In both cases the internal variables are reported. (c) the apo-closed form (named A) and the calcium-bound open form (named B) of Calmodulin (pdb codes: 1WRZ, 1EXR).

## 2.2 Simulation setup and transition path extraction

MD simulations were performed in canonical ensemble using the Langevin (stochastic) thermostat. The timestep was set at 0.01 ps. Simulations had different length, between 20 and 50 ns. The data dumping frequency was on average 0.1 ps$^{-1}$. Simulations were performed with the two different CG FFs (hereafter FF$_A$ and FF$_B$) generated with a bias towards closed and open states (A and B, respectively), and at different temperatures. Simulations were performed with DL_POLY (vs 4.08[21]) and the input was generated with proprietary software.

In order to extract a transition path from the trajectory, we first define the parameter σ based on the root mean square deviation (RMSD$_{A/B}$) of a configuration $\boldsymbol{r} = \{x_i, y_i, z_i\}$ from the reference structures $\boldsymbol{r}^{A/B}$(after alignment[22], to eliminate roto-translations)

$$RMSD_{A/B}(\mathbf{r}) = \sqrt{\frac{1}{N}\sum_i (x_i - x_i^{A/B})^2} \qquad \sigma(\mathbf{r}) = \frac{1}{2}\left(\frac{RMSD_A(\mathbf{r}) - RMSD_B(\mathbf{r})}{RMSD_{A,B}}\right) + \frac{1}{2} \qquad (3)$$

σ ranges between 0 (in A) and 1 (in B). σ is a rough measure of the transition advancement. Clearly, structures with the same σ(**r**) can have different conformations, with different distances from A and B, accounted for by $RMSD_A(\mathbf{r})$ and $RMSD_B(\mathbf{r})$ separately. Therefore, the scatter plot $RMSD_B$ vs $RMSD_A$ will also be considered to have more specific information on the transition path.

| FF term | Functional form | Parameterization | |
|---|---|---|---|
| **Bond** $u_i^b(d_i)$ | restrains | $d_i$ from the reference structure (~3.8Å) | |
| **Bond angle** $u_i^\theta(\theta_i)$ | $\frac{1}{2}k_i^\theta(\cos\theta - \cos\theta_0^i)^2$ | $\theta_0^i$ from the reference structure $k_i^\theta = \frac{B}{\sin^2\theta_0^i}\frac{\sin(\beta\theta_0^i)}{\theta_0^i}$ B = 3000 Kcal/mole β=1.68 rad$^{-1}$ | 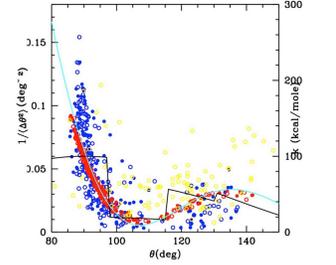 |
| **Dihedral** $u_i^\phi(\phi_i)$ | $A_i^\phi[1 - \cos(\phi - \phi_0^i)]$ | $\phi_0^i$ from the reference structure $A_i^\phi[Kcal/mole] = \begin{cases} 20 & \text{if } \phi_0 \leq 80 \text{ deg} \quad \text{helices} \\ 3 & \text{if } \phi_0 > 80 \text{ deg} \quad \text{strands} \end{cases}$ | |
| **Local** $u_{loc}(r)$ | $\varepsilon^{ij}\left[\left(e^{-\alpha^{ij}(r-r_0^{ij})} - 1\right)^2 - 1\right]$ | $r_{cut}$=8.5Å $\varepsilon^{ij} = 3.0\, e^{-r_{ij}^{0\,8}/6.} + 0.05$ $\alpha^{ij} = 0.6\, e^{-r_{ij}^{0\,8}/6.} + 0.70$ | 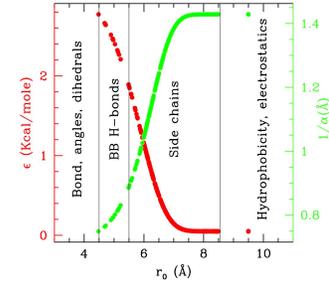 |
| **Non local** $u_{nl}(r)$ | $\varepsilon\left[\left(e^{-\alpha(r-r_0)} - 1\right)^2 - 1\right]$ | $r_0$=9.5Å $\varepsilon$ = 0.05 Kcal/mole $\alpha = 0.70$Å$^{-1}$ | |

**Table 1.** Functional forms (first and second columns) and parameterization (third column) of the MCG FF. An illustration of the statistics-based parameterization procedure is also reported in the plots. Upper plot: The dots represent the inverse bond angle fluctuations as a function of the bond angle, evaluated using atomistic simulations of different test proteins (yellow a globular protein, blue the calmodulin itself, different symbols for different runs). This curve can be fitted as damped sin (cyan line). Assuming statistical equilibrium one has an angle dependent effective elastic constant from the equation $k' = k_B T/\langle\theta^2\rangle$. A further factor $1/\sin^2(\theta_0)$ accounts for the non-exactly harmonic functional form used here (i.e. harmonic cosine) leading to the final functional form for $k_\theta$ reported in the table, which accounts for the secondary structure dependence of the elastic constant (stronger for helices with $\theta_0$~90 deg, softer for strands with $\theta_0$>110 deg). Red dots show the result from a simulation with MGC model with this parameterization. The black line reports the previously used parameter dependence for comparison. For the dihedral term a similar secondary structure dependent parameterization is used, expressed through a simpler step wise dependence on the dihedral value. The non-bonded interactions parameters are reported in the lower plot: dependence of the well depth (ε) and interaction range (1/α) on the equilibrium distance (the shorter the equilibrium distance, the stronger and shorter ranged the interaction). The plot also reports typical interactions included in the corresponding ranges.

In order to identify a limited number of relevant points along the trajectory, we applied the principal path (PP) clustering algorithm[23] to the MD trajectories and extracted reduced trajectories, which retain the salient properties of the original ones. The PP algorithm is a regularized version of the *k*-means clustering algorithm[24], based on the evaluation of a cost functional composed of two parts: the sum of the squared distances of each point from its respective representative structure, and the sum of the squared distances between adjacent representative structures. The relative weight of the two components – the regularization parameter *s* – is obtained by the Bayesian evidence maximization.

The cost functional can be interpreted as an energy, thus the Bayesian posterior probability function is set proportional to the exponential of its negative. The result of the clustering is a "cleaned-up trajectory" of representative structures, used to evaluate σ and energy profiles.

Energies were evaluated both with the CG FFs and at the atomistic level. To this aim, the atomistic structures were rebuilt from the MCG models using Pulchra[25] without any local optimization, then explicitly hydrated and locally optimized using the OPLSe [26] FF with explicit solvent and the Polak-Ribiere conjugate gradient algorithm[27] keeping the backbone frozen during the minimization. The calculations were performed with Schrodinger 2018-2, MacroModel[28].

## 2.2 PROMPT and MAP path search

The PP clustering trajectories are compared with the trajectories obtained from other transition analysis methods. The method MinActionPath[15] (MAP) employs differential equations, obtained by minimizing an action functional including a very simplified potential term representing the protein as a network of harmonic interactions (the elastic network model, ENM)[29]. The equilibrium distances are taken from the reference structures, making the ENM the simplest completely biased model. The solutions to the pair of differential equation are merged by requiring continuity between them. The final result is a single trajectory connecting the two states, reproducing the energy profiles of the mono-stable ENMs near A or B, and with a continuous crossover region.

On the other hand, PROMPT[13] (PRotein cOnformational Motion PredicTion[30]) connects states A and B avoiding relations to any specific FF, by using only structural information. The protein is represented at the CG level and each protein conformation is handled as a set of internal coordinates. The transition path is first guessed e.g. using linear interpolation between extremal configurations $r^A$ and $r^B$. The "admissible motions" are defined, as those preserving all the bond lengths $b_i^J$ and other physical constraints related to bond and dihedral angles ($i$ is the index running along the internal coordinate, and $J$ labels the configuration along the path, from A to B). The path connecting A and B is therefore found by minimizing a kinetic only action integral within the space of admissible motions factorized by rigid roto-translations. The infinite-dimensional variational problem is addressed by discretizing the path between A and B and solved by means of the gradient descent method. The admissible motions are searched by changing the internal free variables of the systems, i.e. $\{\theta_i^J, \phi_i^J\}$ in MCG model; $\theta_i^J$ is treated by interpolation when possible. The detailed description and formal comparison of the three method is reported elsewere[31]. Energies along MAP and PROMPT trajectories were compared using both atomistic (upon rebuilding and side chain optimization as already explained) and MCG FFs.

## 3 Results

### 3.1 Molecular dynamics of the open-closed transition of Calmodulin

Calmodulin (Cam) displays two very different conformations[4], depending on the environmental calcium concentration. The two extremal structures of Cam, i.e. closed (A) and open (B) (see Fig. 1c), correspond to the apo and $Ca^{2+}$-bound state respectively. Because these are, *de facto,* distinct proteins, having different ligands, it is conceptually correct to use two distinct FFs and to perform LD simulation started from A using $FF_B$ to reproduce the A→B transition occurring upon $Ca^{2+}$ binding, and, *vice-versa*, using $FF_A$ for the B→A inverse transition occurring upon $Ca^{2+}$ release. A few data are available for the difference in Gibbs free energy between the folded and denatured proteins ranging between $\Delta G_A$~1.5-3.5[32,33] kcal/mole for the A state and $\Delta G_B$~4.5-6.5 kcal/mole for the B state[33]. Energy alignment is not straightforward, however, one might assume the denatured state as reference, and infer that B state is more stable than A of about 2-4 kcal/mole.

The A-B transition was simulated with LD, in both senses, at 300K (RT) and at 130K. Fig. 2(a,b) reports the energies along the LD simulations. In both cases the transitions are clearly visible in the evolution of σ, passing from 0 to 1 (A→B, green) or from 1 to 0 (B→A, red), though they occur at different times, depending on the simulation parameters and on the FF. In particular, the closed to open transition (green) occurs earlier and more directly, while the inverse open to close transition appears to explore an intermediate conformation with σ~0.4-0.5 for tens of ns before reaching the final state. This is better seen in the RMSD scatter plots reported in panels c and d: the intermediate state, located in the upper right off diagonal part of the plot, persists also after the clustering procedure (joined dots in panels c and d) and is present at high and low temperature, although in the low one it is pushed towards the diagonal. It corresponds to a compact globular conformation, favored over the completely open one by hydrophobicity, but in which the specific contacts of the closed conformation are not formed (see the inset in c and d, red structures). In this work Cam is used only as an example, therefore exploring in detail its transition is out of our scopes. However, we remark that the presence of such mis-folded transition intermediates was previously documented[4]. The intermediate is not visible in the A→B simulations (green), in which the system passes rapidly to B, not even in the PROMPT and MAP trajectories, lying near the diagonal line joining A and B in the RMSD plot. These, additionally, display distorted conformations in the intermediate σ regions. An inspection to the structures with σ~0.5 (reported in purple and cyan in Fig. 2c) shows distortions in the central helix and too contracted terminal regions in the PROMPT structure, and broken chain in the MAP structure.

### 3.2 Data clustering and comparison with PROMPT and MAP

While MAP and PROMT return transition paths made of a few points, the MD simulations explore a large portion of the conformational space returning thousands of conformations. Therefore, in order to compare the methods, we first performed a post-processing and clean-up of the MD trajectories to select a limited number of representative states along it. This can be done in several ways. Fig. 3a reports a simple averaging procedure: the structures are first ordered according to their σ value (red and green dotted/dashed lines), so that A→B transition is read from left to right and B→A from right to left. Once again, the formation of an intermediate cluster at σ=0.4-0.5 is clearly visible in the B→A simulations, beside the large cluster of A type structures and of B type structures in the A→B simulations, respectively. The structures are then grouped according to their σ value in a given number of regular σ intervals; the average energy evaluated in each interval is reported in the plot, for the A→B (green) and B→A (red) simulations at 300K and 130K (dots with error bars). Interesting enough, transitions occur in all cases with a gain of ~20 Kcal/mole (as measured from the starting state, i.e. in each case the opposite of the stable one), irrespective of the temperature and of the FF. As said, comparing the energies resulting from two different FFs is not straightforward. In this case, an inspection of Fig. 2(c,d) shows that the simulation trajectories with $FF_A$ and $FF_B$ get particularly near in a region of the $RMSD_A$-$RMSD_B$ plane corresponding to σ~0.4, indicating that in that area structures belonging to different trajectories are similar. Aligning the energy values for that value of σ in the plot of Fig. 3a generates a small shift leading to B structure more stable than A one of about 3-4 kcal, roughly corresponding to the experimental evaluation. The resulting "activated state structure" corresponds to the intermediate found in the B→A simulations, which turns out to be located ~10 Kcal/mole above the A/B states. This "barrier" value seems rather independent on the simulation temperature, whose effect appears to be a rigid shift of the average energies.

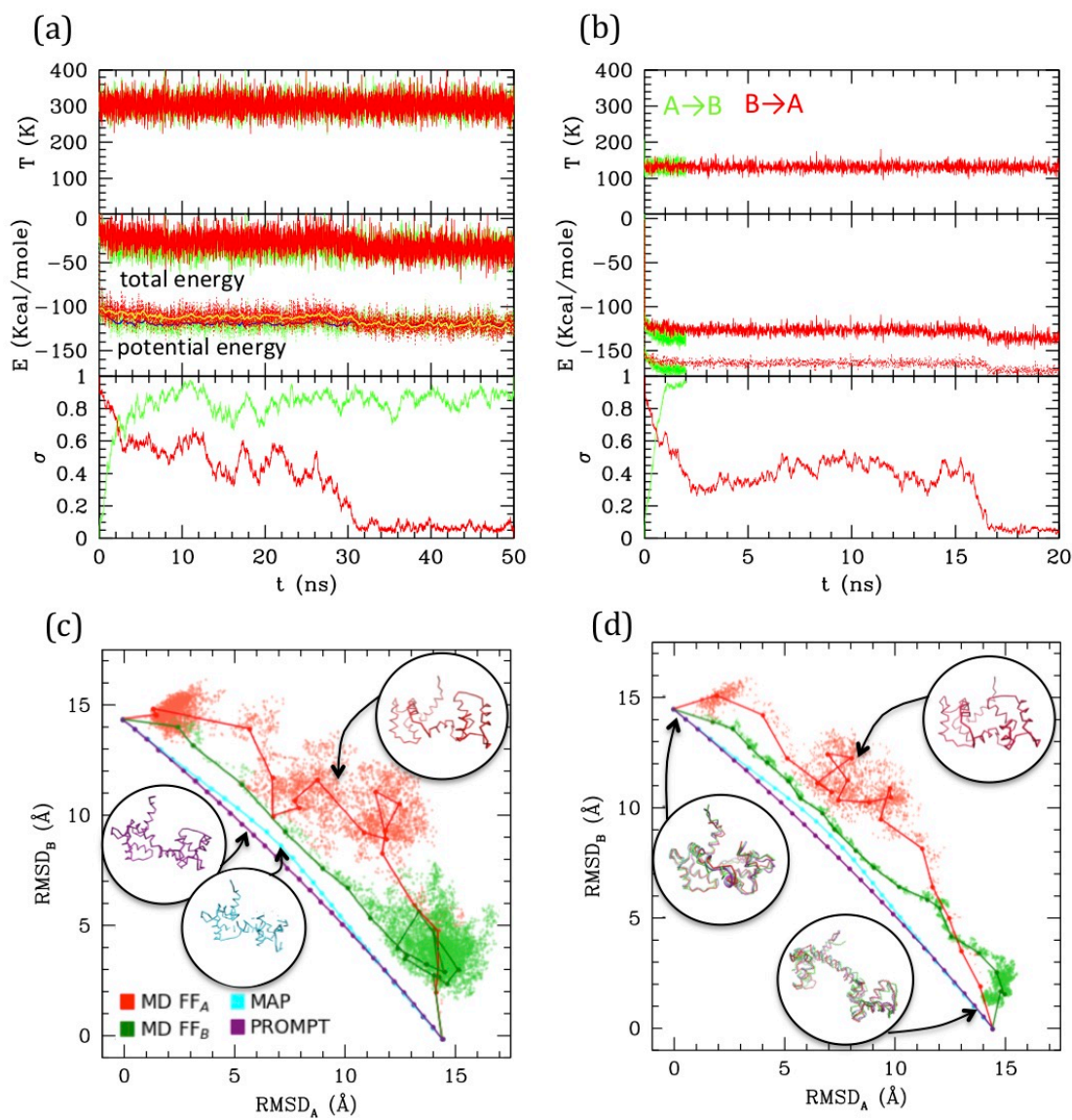

**Fig. 2**. Simulations results from Langevin dynamics at 300K, $\gamma=8ps^{-1}$ (a) and 130K, $\gamma=2ps^{-1}$ (b). Temperature (upper plots), total and potential energies (central plot) and $\sigma$ are reported along the simulations from A to B (using $FF_B$ and starting from configuration A, green lines), and from B to A (using $FF_A$ and starting from configuration B, red lines). For the 300K simulation also the running averages are reported for the potential energy as yellow and blue lines respectively. (c) and (d) Scatter plot of the LD simulations (same color coding as previous) compared with MAP and PROMPT paths evaluation (color coding as in the legend of (c)). The connected dots are the representative elements of the PP clustering procedure. Sample configurations are reported in colors corresponding to the lines and their approximate location in the plots are indicated by arrows.

While the described procedure gives reasonable values of the energies, representative structures along the trajectories are more properly selected via the PP algorithm. This returns a user-defined (20 in this case) number of elements, which are not elements belonging to the trajectories they represent, but rather elements optimizing the structure variance within the trajectory. As a consequence, the energy profiles obtained evaluating the $FF_A$ and $FF_B$ energies onto them (Fig. 3b, solid lines and squared symbols) are rather regular and lie lower in energy with respect to parent trajectories, shown by lines connecting circle symbols (obtained selecting the nearest elements to the optimal ones, filled and empty dots connected by dotted and dashed lines). Remarkably, even after post processing, the main features of the simulation remain: the cluster located at σ~0.4-0.5 is well represented in $FF_A$ simulations, and is located about 10 Kcal/mole above with respect to A and B states.

The optimal element trajectories extracted from the low temperature runs are also reported in Fig. 3c to be compared with the energies evaluated from the MAP and PROMPT trajectories using the MCG FFs. Even after a local optimization, the energies from MAP and PROMPT rapidly increase producing a very large energy barrier at intermediate σ values. An inspection of the structures (reported as insets in the plot) reveals that these arise from severe distortion of the backbone (especially for MAP) and/or steric clashes (both). In particular, the high energy of the intermediate from PROMPT seems to be due to steric clashes in one of the two ends of the protein (highlighted with a yellow circle in Fig. 3c.

Clearly, larger energies on the MAP/PROMPT paths evaluated with MCG FFs are expected, since the low energy path extracted with PP from simulations minimize the MCG Hamiltonian. Therefore, in order to clarify if this energy difference reflects a real larger stability of MCG derived conformations, we rebuilt the atomistic structure of the paths evaluated with all methods and compared their energies evaluated with the atomistic FFs (Fig. 3d), after optimization of the side chain conformation keeping fixed the backbone structure. All methods give comparable energies for structures near A and B states, where in some cases PROMPT and MAP seems to work better than MCG models. However, the atomistic analysis confirms the strong instability of MAP derived structures, displaying unphysical backbone conformation, as shown by the reported Ramachandran plot (upper right inset of Fig. 3d). The instabilities of the PROMPT profile are confirmed in the central σ~0.2-0.8 region, although the Ramachandran plot (central inset) is regular even in there. In fact, in agreement with what found in the MCG model, the instability is not due to a wrong backbone conformation, but to steric clashes in the highlighted area (yellow circle), displaying two sheets whose relative conformation is too close and not correctly aligned.

## 4   Summary and conclusions

In this work we set up a simulation paradigm for finding the transition path of proteins undergoing large conformational transitions, which is a long-standing problem of biophysics. Proteins are modeled by a Cα based coarse-grained representation, while the transition path is explored via classical molecular dynamics simulations with FFs partially biased towards the reference structures. The selection of a representative trajectory among the huge number of configurations explored during molecular dynamics simulations is accomplished by means of the principal path clustering algorithm, which managed to single out trajectories close to those of minimum free energy, yet capable of exploring intermediate states, with a very low computational cost. The comparison with minimal action path and PROMPT can be summarized as follows: MAP returns structures which are reasonable in the near vicinity of the references states, but is unable to provide meaningful ones, even after post-processing, in the intermediate regions. PROMPT returns in addition good backbone local conformations along the whole path, but does not guarantee that amino-acids separated along the chain do not get too near and cause steric clashes, which happens in fact, in the intermediate regions. The MCG simulations, guarantee physically sound structures along the whole path, and can explore also intermediates far from the reference structures, but needs appropriate post-processing and clustering techniques to extract a reaction path. We envision that a synergistic use of these methods might combine accuracy and efficiency in the path search. This possibility is explored in a forthcoming paper[31].

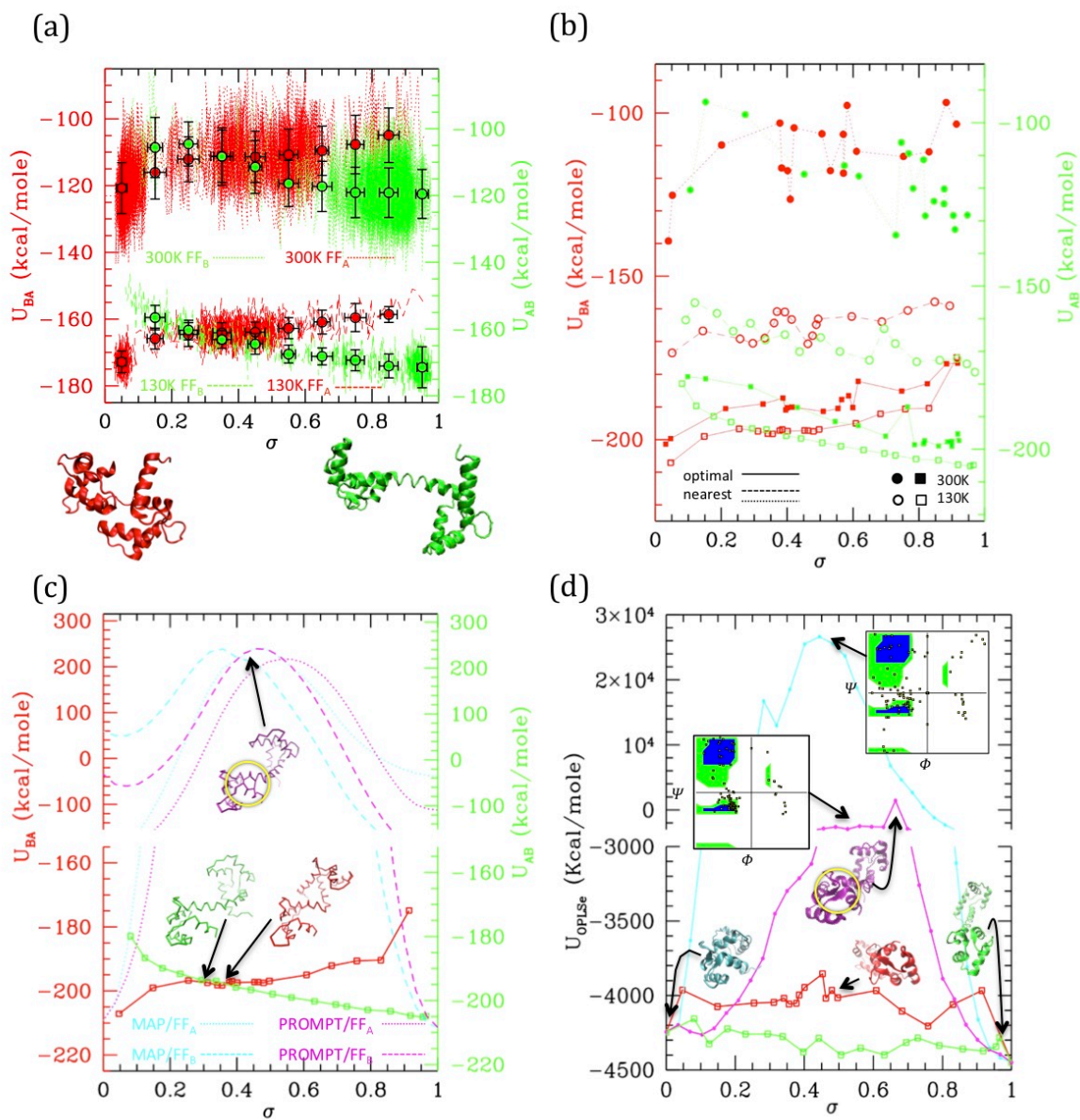

**Fig. 3**. Simulation data analysis and comparison with PROMPT and MAP (a) Potential energy vs σ along the simulations at 300K (dotted lines) and at 130K (dashed lines), with the $FF_A$ (red) and $FF_B$ (green) force fields (scales for $FF_A$ and $FF_B$ are shifted of 3 Kcal/mole to align the activated state as explained in the text. Both scales are reported on the left and right axis, in colors corresponding to the FF they refer to). Colored dot with error bars are averages over subsets of structures classified by σ intervals (errorbars correspond to standard deviations of data from average values). Representative closed (σ=0) and open (σ=1) structures are reported under the plot. (b) Potential energies vs σ evaluated over the representative structures of the clusters outputted by PP procedure. Squares connected by solid lines: representatives optimized by the PP procedure (filled = from the 300K simulation, empty = from the 130K simulations, red with $FF_A$, green with $FF_B$). Circles connected by dashed/dotted lines: same as previous, but evaluated over a trajectory of structures extracted from the simulations, the nearest to the optimal ones. (Same color and empty/filled code as for squares; shift of scales as in (a)). (c) Comparison of the 130K "optimal" energies with energies of trajectories from MAP (cyan) and PROMPT (magenta) evaluated with $FF_A$ (dotted) and $FF_B$ (dashed). Representative structures of the activated states are reported in corresponding colors. Same scale shift as in (a); the vertical scales are broken to zoom over the low energies. (d) Potential energy evaluated with the atomistic FF over the same trajectories as in (c) (same color coding). Representative structures are reported in corresponding colors; the Ramachandran plot of the activated states of PROMPT and MAP are reported (yellow squared dots superimposed to the standard map in colors). Both in (c) and (d) the area with distorted sheets in the activated state of PROMPT is highlighted with a yellow circle.

**Conflict of Interest**

*The authors declare that the research was conducted in the absence of any commercial or financial relationships that could be construed as a potential conflict of interest.*

**Author Contributions**

FD and YP contributed equally to this work. FD have produced data and performed analyses, and participated in writing. YP has produced data and analyses, designed work, contributed ideas and participated in writing. ES and GT have designed the work, contributed ideas, and participated in writing. VT has contributed ideas, designed work and supervised it, and written the paper.

**Funding**

The work of Eugene Stepanov has been supported by the RSF grant 19-71- 30020 "Applications of probabilistic artificial neural generative models to development of digital twin technology for non-linear stochastic systems" (HSE University). The work of Yuri Porozov has been supported by the "Russian Academic Excellence Project 5-100" of Sechenov Medical University.

**Acknowledgments**

We wish to thank Dr. Walter Rocchia for useful discussions and for support in using the software for PP calculations, and Prof. Paolo Carloni, Dr. Giulia Rossetti and Dr. Emiliano Ippoliti for useful discussions on Calmodulin. We also wish to thank Vladimir Kadochnikov for help with Gromacs calculations.

**Supplementary Material**

Numerical raw data (trajectories and energies during simulations, clusters analysis etc) are available as supplementary material.

**Data Availability Statement**

All data used for this work are included in the paper or in supplementary material